\documentclass[prl,showpacs,preprintnumbers,amsmath,amssymb,superscriptaddress,nofootinbib,english,twocolumn]{revtex4}
\usepackage{graphicx}
\usepackage{dcolumn}
\usepackage{bm}
\usepackage{epsfig}
\usepackage{graphicx}
\usepackage{hyperref}
\usepackage[usenames]{color}
\usepackage{url}
\hypersetup{
    colorlinks=true,
    linkcolor=black,
    citecolor=black,
}

\newcommand{\remove}[1]{}

\def\be{\begin{equation}}
\def\ee{\end{equation}}
\def\ba{\begin{eqnarray}}
\def\ea{\end{eqnarray}}

\begin{document}

\title{Modified Gravity Tomography}

\author{Philippe~Brax}
\email[Email address: ]{philippe.brax@cea.fr}
\affiliation{Institut de Physique Theorique, CEA, IPhT, CNRS, URA 2306, F-91191Gif/Yvette Cedex, France}

\author{Anne-Christine~Davis}
\email[Email address: ]{a.c.davis@damtp.cam.ac.uk}
\affiliation{DAMTP, Centre for Mathematical Sciences, University of Cambridge, Wilberforce Road, Cambridge CB3 0WA, UK}

\author{Baojiu~Li}
\email[Email address: ]{baojiu.li@durham.ac.uk}
\affiliation{ICC, Physics Department, University of Durham, South Road, Durham DH1 3LE, UK}
\affiliation{DAMTP, Centre for Mathematical Sciences, University of Cambridge, Wilberforce Road, Cambridge CB3 0WA, UK}
\affiliation{Kavli Institute for Cosmology Cambridge, Madingley Road, Cambridge CB3 0HA, UK}

\date{\today}

\begin{abstract}
We consider  the effect of a canonically normalised scalar field degree of freedom on the dynamics of gravity from small to large scales. We show that the effects of modified gravity can be completely captured by the time variations of the scalar field  mass and its coupling to matter. This leads to  a parameterisation of modified gravity where local constraints are easy to analyse and large scale scale structure effects apparent.
\end{abstract}

\maketitle

Modified gravity is an alternative to dark energy \cite{cst2006}. Plausible scenarios  reproducing  the accelerated cosmic expansion  have been proposed in the past decade, and general parameterisations are currently sought for \cite{zbfs2011}.
A simple possibility uses two functions $\nu(k,a)$ and $\gamma(k,a)$ through the Poisson equation\cite{mu1,mu2} $$-k^2\Phi=4\pi(1+\nu)G_Na^2\delta\rho_m$$ and $$\Psi=(1+\gamma)\Phi.$$ Here $G_N$ is Newton's constant, $\Psi$ and $ \Phi$ are the potentials in the Newtonian gauge: $$ds^2=-a^2(1+2\Psi)d\tau^2+a^2(1-2\Phi)dx^2,$$ where $\tau, x$ are  the conformal time and comoving coordinates. Because $\nu, \gamma$ are free functions of both time ($a$) and space ($k$), such a parameterisation is very general and convenient for phenomenological studies of large scale structure. In the linear regime, it can be shown to be covering several important classes of modified gravity theories, such as $f(R)$ gravity \cite{fR}. These theories differ drastically from the $\nu-\gamma$ parameterisation in the non-linear regime where a screening  of the modified gravity effects is present. Such a shielding effect is apparent in N-body simulations\cite{simu1,simu2}. Moreover, designed for the analysis of the cosmological perturbations in the linear regime, the $\nu-\gamma$ parameterisation is not applicable to other gravitational regimes such as the solar system or laboratory tests of gravity; in particular, it fails to capture the environmental dependence of screened modified gravity models that is so crucial  to evade  local constraints\cite{arXiv:1011.5909}. Also, it focuses on the phenomenological consequences of modified gravity and makes the physics rather obscure, {\it e.g.}, it  gives no hint  about the theoretical properties of the extra degrees of freedom mediating the modification of gravity. Finally, it is not clear that a simultaneous parameterisation of the time and spatial dependence of $\nu, \gamma$ is consistent with any underlying theory. All in all, the $\nu-\gamma$ parameterisation is  useful  for large scale structure in the linear regime  but cannot, in general, lead to a well-defined theory of modified gravity, which must be applicable from small scales where gravity is tested in the laboratory to cosmological scales where non-linear effects are crucial to structure formation.

In this paper we propose a new parameterisation of a broad class of theories which involve a fifth force mediated by a new scalar degree of freedom, such as the chameleon \cite{chameleon}, dilaton \cite{dilaton} and symmetron \cite{symmetron} theories, and  $f(R)$ gravity. The success of these theories relies on certain mechanisms that suppress the fifth force in  local, high matter-density, environments. It may seem  that a full parameterisation of these theories should include not only the temporal and spatial but also the environmental dependences; in fact as we will show below, what we actually need is the temporal dependence of the mass and the coupling to matter alone, which are often  simple power-law functions: once this is given, the behaviour of the theory in different regimes is completely fixed.

The action governing the dynamics of a canonically normalised scalar field $\phi$ in a scalar-tensor theory is
of the general form
\begin{eqnarray}
S &=& \int d^4x\sqrt{-g}\left\{\frac{m_{\rm Pl}^2}{2}{
R}-\frac{1}{2}(\nabla\phi)^2- V(\phi)\right\}\nonumber\\
&& + \int d^4x \sqrt{-\tilde g} {\cal
L}_m(\psi_m^{(i)},\tilde g_{\mu\nu})\,, \label{action}
\end{eqnarray}
where
$g$ is the determinant of the metric $g_{\mu\nu}$, ${ R}$ is
the Ricci scalar and $\psi_m^{(i)}$ are various matter fields
labeled by $i$. A key ingredient of the model is the conformal
coupling of $\phi$ with matter particles. More precisely, the
excitations of each matter field $\psi_m^{(i)}$ follow the
geodesics of a metric $\tilde g_{\mu\nu}$ which is related to the
Einstein-frame metric $g_{\mu\nu}$ by the conformal rescaling
$$\tilde g_{\mu\nu}=A^2(\phi)g_{\mu\nu}.$$
The Klein Gordon equation is modified due to the coupling of the scalar field $\phi$ to matter:
\begin{equation}
\Box \phi= -\beta T + \frac{dV}{d\phi},
\end{equation}
where $T$ is the trace of the energy momentum tensor $T^{\mu\nu}$ and
the coupling of $\phi$ to matter is defined by
\begin{equation}
\beta\equiv m_{\rm Pl}\frac{d\ln A}{d \phi}.
\end{equation}
This is equivalent to the usual Klein-Gordon equation with the effective potential
\be V_{\rm eff}(\phi) = V(\phi) - (A(\phi)-1)T\ee
In the weak-field limit with $$ds^2=-(1+2\Phi_N) dt^2+ (1-2\Phi_N)dx^idx_i,$$ the modified geodesic equation for matter particles reads
\be
\frac{d^2 x^i}{dt^2}= -\partial^i (\Phi_N +\ln A(\phi)).
\ee
where $\Phi_N$ is Newton's potential.
This can be interpreted as the motion of a particle in the effective gravitational potential defined as $$\tilde \Psi= \Phi_N + \ln A(\phi),$$ the scalar field induces a modification of gravity.

We focus on models where the effective potential has a minimum $\phi(a)$ which depends on the scale factor $a$ due to the time variation of the matter density. Using the definition of the scalar mass at the minimum of $V_{\rm eff}$, $$m^2\equiv\frac{\partial V_{\rm eff}(\phi)}{\partial\phi^2}, $$ we deduce the relation
$$
V''\equiv \frac{d^2V}{d\phi^2}= m^2 (a) -  \beta ^2 A(\phi) \frac{\rho}{m_{\rm Pl}^2}-\frac{d \beta}{d\phi} A (\phi) \frac{\rho}{m_{\rm Pl}},
$$
where the couplings to matter $\beta$ can be field dependent and $m^2$ is evaluated at the minimum of $V_{\rm eff}$. Using the minimum equation
\be
\frac{dV}{d\phi}=- \beta A \frac{\rho}{m_{\rm Pl}},
\ee
we find  that the field value at the minimum evolves according to
\be
\frac{d\phi}{dt}=\frac{3H}{m^2} \beta A \frac{\rho}{m_{\rm Pl}}.
\ee
This leads to the solution for the time evolution of the minimum
\begin{equation}
\phi(a)=  \frac{3}{m_{\rm Pl}}\int_{a_{\rm ini}}^a \frac{\beta (a)}{a m^2(a)}\rho (a)  da +\phi_c,\label{phi}
\end{equation}
where $\rho$ is the density of non-relativistic matter species, $\phi_c$ the initial value of the scalar field and we have taken $A(\phi)\approx1$ as the temporal variation of fermion masses must be very weak (see later). If the coupling $\beta$ is expressed in terms of the field $\phi$ and not the scale factor $a$, this is also equivalent to
\be
\int_{\phi_c}^\phi \frac{d\phi}{\beta(\phi)}=  \frac{3}{m_{\rm Pl}}\int_{a_{\rm ini}}^a \frac{1}{a m^2(a)}\rho (a) da.
\ee
Similarly the minimum equation implies that the value of the potential at the minimum is given by
\begin{equation}
V(a)=V_0 -3  \int_{a_{\rm ini}}^a \frac{\beta(a)^2}{am^2(a)} \frac{\rho^2}{m^2_{\rm Pl}} da.
\label{V}
\end{equation}
It turns out that (\ref{V}) and (\ref{phi}) define a parametric representation of the potential $V(\phi)$ and the coupling $\beta (\phi)$. This implicit dependence is valid for all the values of $\phi(a)$ as $a$ varies. In practice, this allows one to reconstruct the full dynamics of the model for  $\phi$ ranging from its value after inflation to now.
This defines the bare scalar field potential $V(\phi)$ parametrically when $\beta (a)$ and $m(a)$ are given and allows one to reconstruct the full dynamics of the models defined by the action (\ref{action}).

As a first example, let us consider the important case of a non-vanishing coupling function $\beta(a)$. Defining $\rho= \frac{\rho_0}{a^3}$, $\beta (a)= \beta_0 g(a)$  and $m= m_0 f(a)$, in which a subscript $_0$ denotes the present-day value, we find that
\begin{equation}\label{eq:phi_chameleon}
\frac{\phi -\phi_c}{m_{\rm Pl}}=9 \beta_0\Omega_{m0}\frac{H_0^2}{m^2_0} \int_{a_{\rm ini}}^a da\frac{g(a)}{a^4 f^2(a)},
\end{equation}
where $H_0$ is the Hubble constant and $\Omega_m^0=\rho_0/3H_0^2m^2_{\rm Pl}$. Let us specialise to the case where
\be
m(a)=m_0 a^{-r},\ \beta(a)= \beta_0
\ee
where $r>3/2$.
We find that
\be
\frac{\phi(a)-\phi_c}{m_{\rm Pl}}= \frac{9\beta_0 \Omega_{m0} H_0^2}{(2r-3) m_0^2} (a^{2r-3}-a_{\rm ini}^{2r-3})
\ee
and
\be
V(a)= V_0 -\frac{27 \beta_0 \Omega_{m0}^2 m_{\rm Pl}^2 H_0^4}{(2r-6) m_0^2}(a^{2r-6}-a_{\rm ini}^{2r-6})
\ee
from which we deduce that
\begin{eqnarray}
V(\phi)&=& V_0-\frac{27a_{\rm ini}^{2r-6} \beta_0 \Omega_{m0}^2 m_{\rm Pl}^2 H_0^4}{(2r-6) m_0^2}\nonumber\\ &\times& ((1+\frac{(2r-3) m_0^2}{9a_{\rm ini}^{2r-3}\beta_0 \Omega_{m0} H_0^2} (\frac{\phi-\phi_c}{m_{\rm Pl}}))^{2(r-3)/(2r-3)} -1)\nonumber \\
\end{eqnarray}
which defines inverse power law chameleon models when $3/2<r<3$ \cite{Brax:2011ta} and power law models when $r>3$. In the latter case and when $\beta_0=1/\sqrt{6}$, these models are equivalent to large curvature $f(R)$ gravity. Some choices of such $f(R)$ models that lead to viable cosmologies are summarised in Table~I, and we see that the different $f(R)$ models in \cite{hs2007,s2007,ftbm2007,bbds2008,lb2007}  result in a power-law evolution of $m$ with $r\ge 3$. This can be easily seen by choosing
\be
f(R)=R- \frac{2\Lambda_0^4}{m_{\rm Pl}^2} -\frac{f_{R_0}}{n}\frac{R_0^{n+1}}{R^n}
\ee
expanding in $R_0/R$ for $R\gtrsim R_0$ where $R_0$ is the curvature now.
Using the equivalence
\be
f_R= e^{-2 \beta_0 \phi_R/m_{\rm Pl}}
\ee
and
\be
V(\phi_R)= \frac{m_{\rm Pl}^2}{2} \frac{Rf_R -f(R)}{f_R^2}
\ee
we find that
\be
\frac{R}{R_0} \approx (-\frac{2\beta_0 \phi_R}{m_{\rm Pl} f_{R_0}})^{1/(n+1)}
\ee
and
\be
V(\phi_R)= \Lambda_0^4 + \frac{n+1}{n} {f_{R_0}}R_0 (-\frac{2\beta_0 \phi_R}{m_{\rm Pl} f_{R_0}})^{n/(n+1)}
\ee
which is a power law model with $n= \frac{2}{3}r -2$.
It can also be obtained that the mass $m_0$ is related to $f_{R_0}$ according to
\be
\frac{m_0}{H_0} = \sqrt{\frac{4\Omega_\Lambda + \Omega_{m0}}{(n+1)\vert f_{R_0}\vert}}
\ee
This completes the identification of the large curvature models with the $m(a)-\beta(a)$ parameterisation.
\begin{table}
\label{tab:fr_tab}
\caption{Some $f(R)$ gravity models,  the corresponding $V(\phi)$ in the Einstein frame, and the evolution of the scalar field mass $m(a)$. $\zeta, n, w, \Lambda, R_{\ast}$ are constant parameters, with $n>0, 0<w\ll1$.}
\begin{tabular}{@{}lccc}
\hline\hline
Ref. & \ Asymptotic $f(R)$\  & \ \ \ Equivalent $V(\phi)$\ \ \ & $m(a)/m_0$\\
\hline
\cite{hs2007,s2007,ftbm2007,bbds2008} & $-2\Lambda+\left(\frac{R_\ast}{R}\right)^n$ & $\zeta\left[1-\exp(-\beta\kappa_4\phi)\right]^{\frac{n}{n+1}}$ & $a^{-\frac{3}{2}(n+2)}$\\
\cite{bbds2008} & $-2\Lambda-\xi\ln\left(\frac{R}{R_\ast}\right)$ & $\Lambda-\zeta\ln(\kappa_4\phi)$ & $a^{-3}$\\
\cite{lb2007} & $-2\Lambda\left(\frac{R}{R_\ast}\right)^w$ & $\frac{\zeta\exp(\beta\kappa_4\phi)}{\left[1-\exp(-\beta\kappa_4\phi)\right]^{\frac{w}{1-w}}}$ & $p+qa^{-3}$\\
\hline\hline
\end{tabular}
\end{table}

As another example, consider a very different, so-called dilaton, model in which the coupling function $\beta(\phi)$ vanishes for a certain value $\phi_\ast$ of the scalar field $\phi$. It is enough to study the dynamics in the vicinity of the field $\phi_\ast$, where $$\beta (\phi) \approx A_2 (\phi-\phi_\ast),$$ from which we deduce that
$$
\ln\left\vert\frac{\phi-\phi_\ast}{\phi_c-\phi_\ast}\right\vert = 9 A_2\Omega_{m0}H_0^2 \int_{a_{\rm ini}}^a \frac{da}{a^4 m^2 (a)},
$$
and therefore
\begin{equation}
\vert \beta (\phi)\vert = \vert \beta (\phi_c)\vert  e^{9 A_2\Omega_{m0}H_0^2 \int_{a_{\rm ini}}^a \frac{da}{a^4 m^2 (a)}}.
\label{eq:beta}
\end{equation}
This expression will be useful to analyse gravitational tests of dilaton models.

The $m(a)-\beta(a)$ parameterisation allows one to study all the different regimes of the models. Let us first consider the background cosmology.
Constraints from the creation of the light elements imply that the variation of fermion masses during Big-Bang Nucleosynthesis (BBN) must be small. This implies that the scalar field must have settled at the minimum $\phi(a)$ of the effective potential before the electron decoupling\cite{Brax:2004qh}. As long as the mass of the scalar field $m(a)$ is much greater than the Hubble rate, $m(a)  \gg H(a) $, the minimum of the effective potential $V_{\rm eff}$ is stable.
Indeed if this were not the case then the scalar field would receive a kick due to the abrupt variation of the trace of the energy momentum during the transition and induce an ${\cal O}(1)$ change in the fermion masses. This is not the case if the field is at the minimum as the steepness of the potential, there $m\gg H$, prevents large excursions
of the scalar field from the minimum. At the transition $\rho \sim m_e^4$  and $z_e\sim 5\cdot 10^8$,
the matter density at the  electron decoupling  is of order $10^{-5}{\rm g\cdot cm^{-3}}$. We must impose  that the field has settled at the minimum of the effective potential earlier than the electron decoupling for instance with $z_{\rm ini}\sim 10^{2} z_e$. At this particular redshift, the energy density of matter is of the order of $10{\rm g\cdot cm^{-3}}$. Hence the initial value $\phi_c$ of the scalar field corresponds to the scalar field value in ordinary matter such as on the Earth, the Sun or laboratory test masses.
Given the time evolutions of the mass $m(a)$ and coupling $\beta (a)$, one can reconstruct the dynamics of the scalar field $\phi(a)$ for densities ranging from cosmological to solar system values using Eq.~(\ref{phi}). By the same token, the interaction potential can be reconstructed for all values of $\phi$ (or $\rho$) of interest using Eq.~(\ref{V}), from the solar system and Earth to cosmological background: a tomography of modified gravity.

The $m(a)-\beta(a)$ parameterisation can also be used to analyse the gravitational tests in the solar system.
For the chameleon models, i.e. the first example, and evaluating Eq.~(\ref{eq:phi_chameleon}) in the galactic vacuum, we find that
$$
\frac{\phi_G -\phi_c}{m_{\rm Pl}}= 9 \beta_0\Omega_{m0}\frac{H_0^2}{m^2_0} \int_{ a_{\rm ini}}^{a_G} da \frac{g(a)}{a^4 f^2(a)},
$$
where $a_G\approx 10^{-2}$ is the scale factor when the matter density in the cosmological background equals the galactic density $\rho_G\approx 10^6 \rho_c$. Here
$\phi_G$ is the value of the scalar field in the galaxy and $\phi_c$ the field inside ordinary matter.
Local tests are satisfied when the thin shell effect is at play where we define the thin shell factor
$$\frac{\Delta R}{R}=\frac{\phi_G -\phi_c}{6m_{\rm Pl}\beta_c \Phi_\odot},$$ and   the modification of gravity in the solar system felt by a satellite such as the Cassini probe has a strength $2\beta_G\beta_c\frac{3\Delta R}{R}$ in which $\beta_G$ is the coupling in the galactic vacuum,  $\Phi_\odot$ is the value of the solar Newtonian potential ($\Phi_\odot\sim 10^{-6}$) and $\beta_c$ is the coupling inside a dense body. The overall result should be less than $10^{-5}$ to comply with the Cassini bound in the solar system\cite{Bertotti:2003rm}. This condition is independent of $\beta_c$ and reads
$$
\beta_0\beta_G \int_{ a_{\rm ini}}^{a_G} da\frac{g(a)}{a^4 f^2(a)}\lesssim 10^{-5} \frac{m_0^2}{9\Omega_{m0} H_0^2} \Phi_\odot.
$$
The integral $I\equiv\int_{a_{\rm ini}}^{a_G} da \frac{g(a)}{a^4 f^2(a)}$ is potentially divergent for small values of $a_{\rm ini}\sim10^{-10}$. Hence we must impose that $f(a)^2/g(a)$ compensates the $1/a^4$ divergence in the integrand. Typically,  we can parameterise $$f(a)= a^{-r},\ g(a)=a^{-s}.$$ We must then impose $2r-s>3$, which leads to
$
I\approx \frac{a_G^{2r-s-3}}{2r-s-3}.
$
The case $s=0$ and $r>3/2$ has already be seen to correspond to inverse power law chameleons and large curvature $f(R)$ models.
Adopting $\beta_G\approx 10^{2s} \beta_0$, we find that there is an interplay between $\beta_0$ and $m_0/H_0$: $$\frac{\beta_0^2 H_0^2}{m_0^2} \lesssim 10^{4r-4s-12}(2r-s -3) \frac{\Phi_\odot}{\Omega_{m0}}.$$ In the above we have assumed that galaxies have a thin shell to minimise the disruption of their dynamics, although the necessity of this condition should be ascertained using N-body simulations \cite{hs2007}. Enforcing the thin-shell condition imposes $$\vert \phi_G -\phi_0\vert \lesssim 6 \beta_0 \Phi_G$$ where the galactic Newtonian potential is $\Phi_G\sim 10^{-6}$ and
$$\frac{\phi_0 -\phi_G}{m_{\rm Pl}}= 9 \beta_0\Omega_{m0}\frac{H_0^2}{m^2_0} \int_{ a_{G}}^{1} da \frac{g(a)}{a^4 f^2(a)}$$ implying that \be\frac{m_0}{H_0} \gtrsim 10^{3}\ee
where $\phi_0$ is the cosmological value of $\phi$ now.
This condition is independent of $\beta_0$ and means that any screened modified gravity model will have effects on Mpc scales only.

Strong constraints can also be obtained from laboratory experiments. Using the fact that the initial matter density at $z_{\rm ini} \sim 10^{10}$ corresponds to the matter density inside laboratory test bodies, we have $m_{\rm lab} \sim 10^{10r} m_0$. Gravity is not modified provided tests bodies have a thin shell, $$\vert \phi_{\rm lab}-\phi_c\vert \lesssim 6\beta_{\rm lab} m_{\rm Pl}\Phi_{\rm lab}$$ where $\Phi_{\rm lab}\sim 10^{-27}$ for typical test bodies in cavity experiments of size $R$, $\phi_{\rm lab}= \phi (a_{\rm lab})$ where $m(a_{\rm lab}) \sim 1/R$ . A weaker condition is that $m_{\rm lab} d \gg 1$ where $d$ is the size of the test body implying that $m_{0} d \gg 10^{-10r}$. Both constraints can be found in figure 1.
Finally, the scalar field  mass is larger than the Hubble rate since $a=a_{\rm ini}$ provided $r\gtrsim 2 - \ln\left(\frac{H_0}{m_0}\right)/\ln a_{\rm ini}$.

For the dilaton models,  if the coupling now is of order unity, according to Eq.~(\ref{eq:beta}),  $\beta(\phi_c)\lesssim 10^{-2.5}$  can be achieved provided that $A_2 >0$ and that the time variation of $m(a)$ is slow and does not compensate the $1/a^4$ divergence in the integrand. In this situation, the coupling function $\beta$ converges exponentially fast towards zero: this is the Damour-Polyakov mechanism \cite{dp1994}. Alternatively, a  smooth variation of the coupling to matter and therefore interesting consequences for large scale structures are  achieved when the mass of the scalar field compensates exactly the $1/a^4$. This is obtained for models with $m^2 (a) = 3A_2 H^2 (a)$ where $A_2 \gg 1 $ here. Indeed,  $H(a) \sim 1/a^2$ in the radiation era implying that the time variation of $\beta$ until the matter-radiation equality is very small. In the matter era $H(a)\sim 1/ a^{3/2}$ implying a power law variation of $\beta$ with $a$. These models have been extensively studied and correspond to the environmentally dependent dilatons \cite{dilaton} where effects of modified gravity are also only effective at the Mpc scale.

\begin{figure}
\begin{center}
\includegraphics[scale=0.3]{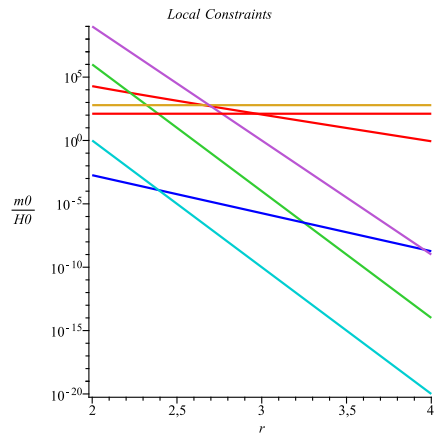}
\caption{The constraints on $m_0/H_0$ as a function of $r$ for $\beta_0=1/\sqrt{6}$ and $s=0$. Valid models must be above the red (solar system), mauve (cavity) green ($m>H$), light blue ($md \gtrsim 1$), light red ($\dot \mu$) and brown (galaxy) lines. The blue line gives the detectability of effects on the CMB by the  Planck satellite. The strongest constraints are the cavity and galactic bounds for small and large $r$ respectively. Models with $r\gtrsim 3$ satisfy the constraints and can lead to a modified gravity regime on large scales. }
\end{center}
\end{figure}

The local constraints give that $m_0/H_0\gtrsim 10^3$ which implies that astrophysical effects of modified gravity can only occur on Mpc scales. This fact can be easily seen by stydying the growth of structures in the linear regime.
In the matter dominated era, the density contrast of cold dark matter (CDM) evolves according to
\be
\delta_c'' +{\cal H} \delta_c' -\frac{3}{2} {\cal H}^2 \frac{\rho_c\delta_c}{\rho_c +\rho_\gamma +\rho_b}\left(1+ \frac{2\beta^2 }{1+\frac{m^2a^2}{k^2}}\right)=0.
\ee
Hence inside the Compton radius $k \gtrsim am(a)$, perturbations grow anomalously. Modified gravity can be effective if in the recent past of the Universe the Compton radius is of the order of 10 Mpc, i.e. $m_0/H_0 \gtrsim 10^3$. We have seen that this is a consequence of local constraints for screened modified gravity models. On the other hand, on such scales non-linear effects cannot be neglected. The strength of the $m(a)-\beta(a)$ parameterisation is that the non-linear regime can be easily described.
Matter clustering on galactic and cluster scales is an important probe of modified gravity. On Mpc scales, the nonlinearity in both the structure formation process and the dynamics of the scalar field requires full numerical simulations \cite{lb2011,lztk2011}. The $\beta(a), m(a)$ parameterisation can completely specify the nonlinear dynamics of $\phi$. To see this, note that in the quasi-static limit the scalar field is governed by
\be
\vec{\nabla}^2\phi = -(\alpha_\phi\rho_m-\bar{\alpha}_\phi\bar{\rho}_m)+V_{,\phi}(\phi)-V_{,\phi}(\bar{\phi}),
\ee
where the overbar means the background value. With $\phi(a)$ reconstructed from $\beta(a)$ and $m(a)$, and $V_{,\phi}(a)$ from $\beta(a)$ and $\rho(a)$, one can easily obtain $V_{,\phi}(\phi)$ analytically or numerically, and this can be used to solve the quasi-static dynamics  numerically. An advantage is that {\it temporal} functions $m(a), \beta(a)$ completely specify the dynamics of $\phi$, in particular its {\it spatial} configuration, and there is no need for a  $k$-space parametrisation.

On the other hand, linear scales could also be of interest for deciphering modifying gravity if, for scales entering the horizon before radiation-matter equality, the anomalous growth of perturbations  plays a role on the peak structure of the cosmic microwave background (CMB). This  happens provided the ratio $$\tilde\beta= \frac{\beta}{(1+\frac{m^2a^2}{k^2})^{1/2}}$$ is of order one \cite{Brax:2011ta}. The scalar field mass at last scattering  is given by $a_{\rm CMB} m_{\rm CMB}\approx  10^{3r-3} m_0$, hence we find that
$$
\tilde \beta_{\rm CMB} \approx \frac{k_{\rm CMB}\beta_{\rm CMB}}{a_{\rm CMB} m_{\rm CMB}} \approx 10^{3-3r+3s}\frac{k_{\rm CMB}}{H_0}\frac{\beta_0 H_0}{m_0},
$$
where $k_{\rm CMB}$ characterises the scale of the horizon at the last scattering. The modified gravity effects can be seen on the CMB provided $10^2\lesssim \omega\approx1/(2\tilde{\beta}^2_{\rm CMB}) \lesssim 3\cdot 10^3$ \cite{ck1999}, giving constraints on the model parameters through
$$
\frac{\beta_0 H_0}{m_0}\approx \frac{1}{\sqrt{2\omega}} \frac{H_0}{k_{\rm CMB}} 10^{3(-1+r-s)},
$$
in which typically we have $H_0/k_{\rm CMB}\sim\mathcal{O}(0.1)$. The previous condition on the observability of modified gravity by the Planck satellite is compatible with the solar system constraint  provided $r-s\lesssim -7$  and therefore $\beta_0 H_0 /m_0 \lesssim 10^{-23}$. This implies that for reasonable values of $\beta_0$, $m_0/H_0$ would be so large that no effect of modified gravity on large scale structures (LSS) would be present. Hence, for models with a power law dependence of both the mass and the coupling to matter, effects on both the CMB and LSS are not compatible.

The scalar field  also has an effect  on gauge couplings and particle masses. The fermion masses are given by $m_F(\phi)=A(\phi) m_F^0$
where $m_F^0$ is the bare mass in the Lagrangian. Meanwhile, quantum effects such as the presence of heavy fermions lead to the coupling of $\phi$ to photons $$S_{\rm gauge}= -\frac{1}{4g^2} \int {\rm d}^4 x \sqrt{-g} B_F(\phi) F_{\mu\nu} F^{\mu\nu},$$ where $g$ is the bare coupling constant and \be B_F(\phi)= 1 + \beta_\gamma \frac{ \phi}{m_{\rm Pl}} + \dots.\ee Depending on the model, the coefficients $\beta$ and $\beta_\gamma$ can be related. Here we will consider them to be free parameters.

The scalar coupling to the electromagnetic field could lead to a dependence of the fine structure constant on $\phi$ as \be\frac{1}{\alpha}= \frac{1}{\alpha_0} B_F (\phi),\ee
implying that \be\frac{\dot \alpha}{\alpha}\approx -\beta_\gamma \kappa_4 \dot{\phi}.\ee Using the evolution equation we find that \be\frac{\dot \alpha}{H \alpha}\approx -9\beta_\gamma \beta \Omega_m \frac{H^2}{m^2}.\ee
Hence the negative variation of the fine structure constant in one Hubble time is related to the small ratio $\frac{H}{m} \ll 1$ and the couplings of $\phi$ to matter and photons.
The best experimental bound on the variation of $\alpha$ now comes from Aluminium and Mercury single-ion clocks: $\frac{\dot \alpha}{\alpha}{\large\vert}_0 = (-1.6\pm 2.3)\cdot10^{-17}{\rm yr}^{-1}$. Taking $H_0^{-1} \sim 1.5\cdot10^{10} {\rm yr}$, we get the conservative bound $$\left\vert\frac{\dot \alpha}{H\alpha}\right\vert_0  \lesssim 2\cdot 10^{-7}.$$ As a result, the experimental bounds on the time variation of $\alpha$ lead to constraints on $\beta_0\beta_{\gamma 0}$ as $\beta_0 \beta_{\gamma 0}\lesssim 0.8\cdot10^{-7} \frac{m_0^2}{H_0^2}$. For models with $\beta_0={\cal O}(1)$, $\Omega_m\sim 0.25$ and $m_0/H_0\approx 10^{3}$ where effects on LSS are present, this is a tighter bound than present experimental ones\cite{chase} \be \beta_{\gamma 0} \lesssim 0.1 \ee

Fundamental fermions such as the electrons have a universal mass dependence $m_F= A(\phi) m_F^0$, implying that \be\frac{\dot m_F}{H m_F} = 9\beta^2 \Omega_m \frac{H^2}{m^2}.\ee Similarly, nucleons such as the proton have a mass given by the phenomenological formula
$$
m_p=C_{\rm QCD} \Lambda_{\rm QCD} + b_u m_u + b_d m_d + C_p \alpha,
$$
where $\Lambda_{\rm QCD}$ is the QCD scale, $b_u+ b_d \sim 6$, $b_u-b_d \sim 0.5$, $C_{\rm QCD}\sim 5.2$, $m_u^0\sim 5 {\rm MeV}$, $ m_d^0\sim 10 {\rm MeV}$ and $C_p\alpha_0 \sim 0.62 {\rm MeV}$.
Of course the main source of uncertainty here follows from the lack of our knowledge about the coupling of scalars to gluon. This leads to weaker bounds than the ones coming from the time variations of $\alpha$. Assuming conservatively that $\Lambda_{\rm QCD}$ is scalar independent\footnote{If quantum effects  lead to a coupling of $\phi$ to gluons with $B(\phi)\approx1+\beta_g\kappa_4\phi$ like for photons, then  $\Lambda_{\rm QCD}\propto(1+\beta_g\kappa_4\phi)^{-\frac{2}{27}}\exp\left[-\frac{2\pi}{9\alpha_{\rm S}(M_Z)}\beta_g\kappa_4\phi\right]$ \cite{lc2006}, where $\alpha_{\rm S}(M_Z)\approx0.12$ is the running strong coupling at the energy scale of the weak $Z$ boson mass. The time variation of $\Lambda_{\rm QCD}$ can be much faster than that of $\alpha$, leading to stronger constraints. For a recent analysis, see \cite{Luo:2011cf}}, we get
\be
\frac{\dot m_p}{H m_p} \approx 9\Omega_m \beta \frac{H^2}{m^2}\left ( \frac{b_u m_u^0 + b_d m_d^0}{m_p}\beta - \frac{C_p \alpha_0}{m_p} \beta_\gamma\right).
\ee
It is particularly important to study the variation of $\mu= \frac{m_e}{m_p}$ from which we find that its time variation  is positive for modified gravity models: \be\frac{\dot \mu}{\mu}\approx 9\Omega_m \beta \frac{H^2}{m^2}\left ( \beta+ \frac{C_p \alpha_0}{m_p} \beta_\gamma\right).\ee The current experimental constraint is $\frac{\dot \mu}{\mu}\large\vert_0 = (-3.8\pm 5.6) 10^{-14}{\rm yr}^{-1}$
which yields the upper bound on $\beta_0$: \be\beta_0^2 \lesssim 10^{-5}\frac{m_0^2}{H_0^2}.\ee For $\beta_0={\cal O}(1)$, this entails that $m_0/H_0\gtrsim 10^{2.5}$. In  Figure 1, we summarise all the constraints when $s=0$ and $\beta_0=1/\sqrt{6}$ (corresponding to inverse power law chameleons and large curvature $f(R)$ models) and show that models with $r\gtrsim 3$ (large curvature $f(R)$ models) are compatible with effects of modified gravity on LSS.

Modified gravity models  compatible with local  experiments must inherit some mechanism to suppress a potential fifth force. This makes these models very nonlinear. We have shown that a large class of such models can be {\it fully} parameterised by only two {\it temporal} functions: the scalar field mass and the coupling to matter allowing one to reconstruct the full action using (\ref{phi}) and (\ref{V}). An  important implication is that instead of studying individual models, one can focus on the effects entailed by the choice of  these two functions. We have shown that many known models can be reconstructed by using a simple power-law form for these  two functions.
Solar system tests, laboratory experiments  and the variation of fundamental couplings and masses can be easily studied and lead to strong constraints on models. In cosmology, the CMB is rather weakly influenced by the fifth force, while on Mpc scales the LSS can be largely affected. On such scales the nonlinearity of both the structure formation and the models is important, and $N$-body simulations are needed to fully understand the model behaviour. This work is ongoing.

\begin{acknowledgments}
We would like to thank  Justin Khoury and Jean-Philippe Uzan for useful discussions.
\end{acknowledgments}

\end{document}